\documentclass[pra,aps,10pt,twocolumn,superscriptaddress,showpacs,amsmath,amssymb,amsfonts]{revtex4-1}
\usepackage{bm}
\usepackage{graphicx}
\usepackage{subfigure}
\usepackage{array}
\usepackage[usenames,dvipsnames]{color}
\usepackage{natbib}
\usepackage[breaklinks=true,pdfborder={0 0 0},colorlinks=true,linkcolor=MidnightBlue,citecolor=MidnightBlue,urlcolor=MidnightBlue]{hyperref}
\newcommand{\mbf}[1]{{\textbf #1}}
\newcommand{\aver}[1]{\left\langle #1 \right\rangle}

\begin{document}

\title{Soliton Molecules In Dipolar Bose-Einstein Condensates}
\author{Kazimierz {\L}akomy}
\affiliation{Institut f\"ur Theoretische Physik , Leibniz Universit\"at, 
Hannover, Appelstrasse 2, D-30167, Hannover, Germany}
\author{Rejish Nath}
\affiliation{Max Planck Institute for the Physics of Complex Systems, 
N{\"o}thnitzer Strasse 38, D-01187 Dresden, Germany}
\affiliation{IQOQI and Institute for Theoretical Physics, 
University of Innsbruck, A-6020 Innsbruck, Austria}
\author{Luis Santos}
\affiliation{Institut f\"ur Theoretische Physik , Leibniz Universit\"at, 
Hannover, Appelstrasse 2, D-30167, Hannover, Germany}
\date{\today}

\begin{abstract}
Dipolar interactions support the formation of inter-site soliton molecules in 
a stack of quasi-1D traps. We show that the stability and properties of 
individual solitons, and soliton molecules in such a geometry crucially depend on 
the interplay between contact and dipolar interactions. In particular, two 
different quasi-1D soliton regimes are possible: a 1D soliton characterized 
by purely repulsive DDI and a 3D soliton for which a sufficiently large dipole 
moment renders the DDI attractive. Furthermore, we find that contrary to the case 
of dimers of polar molecules, the soliton dimers exhibit a nontrivial behavior 
of the elementary excitations that stems from the competition between on-site 
and inter-site DDI. Finally, we prove the existence of soliton trimers in a regime 
where molecular trimers do not occur. We demonstrate that the soliton molecules 
that we report are well feasible under realistic experimental conditions.

\end{abstract}
\maketitle

\section{Introduction}

Recent developments in experiments on ultra-cold polar molecules~\cite{Weidemuller:2008,
Ni:2008,Ospelkaus:2010}, atoms with large magnetic dipole moment~\cite{Pfau:2005,
Beaufils:2008,Lev:2011,Ferlaino:2012}, and Rydberg atoms~\cite{Tong:2004} opens new 
promising perspectives in the rapidly progressing research on dipolar quantum gases. 
Interestingly, the presence of long-range and anisotropic dipole-dipole interactions 
(DDI) essentially modifies the behavior of quantum gases leading to a wealth of a new 
physics~\cite{Baranov:2008,Lahaye:2009}.

Dipolar effects are particularly relevant to what concerns the nonlinear properties of 
dipolar Bose-Einstein condensates (BECs). Crucially, whereas non-dipolar BECs present a 
local Kerr-like type of nonlinearity, the nonlinearity in dipolar BECs exhibits a nonlocal 
character, similar to that in plasmas~\cite{Litvak:1975}, photorefractive media 
\cite{Segev:1992, Duree:1993} and nematic liquid crystals~\cite{Peccianti:2002, Peccianti:2004}. 
Interestingly, this nonlocality results in novel physical phenomena, including 
stabilization of two-dimensional solitons~\cite{Pedri:2005, Tikhonenkov:2008}.

The long-range character of the DDI plays a substantial role in the physics of dipolar 
gases in optical lattices, even in the absence of inter-site hopping. While a 
non-dipolar gas in such a deep lattice may be considered as a system of mutually independent 
gases, nonlocal inter-site interactions in a dipolar gas couple the disjoint sites. 
In particular, in the physics of polar molecules this feature gives rise to a variety of 
unprecedented few-body bound states such as inter-site dimers~\cite{Pikovski:2010,Potter:2010}, 
trimers~\cite{Wunsch:2011,Dalmonte:2011} and filaments~\cite{Wang:2006,Klawunn:2010}.

The inter-site interactions play also a key role in the behavior of a dipolar 
condensate in an optical lattice. Specifically, they have been found to fundamentally 
modify the BEC excitation spectrum~\cite{Klawunn:2009,Wunner:2009} and to affect significantly 
the stability of the condensate~\cite{Wilson:2011,Mueller:2011}. Moreover, inter-site 
interactions may lead to a correlated modulational instability, in which a locked 
density modulation pattern is shared among non-overlapping sites, after a quench of 
a condensate into instability. Interestingly, such correlated modulational instability 
may result in the dynamical formation of soliton filaments and crystalline 
structures~\cite{Lakomy:2012}.

In this paper we analyze in detail the physics of dipolar bright solitons in a stack of 
quasi-1D condensates. We focus on the stability and properties of soliton dimers and 
trimers, which constitute the building blocks of the above-mentioned soliton filaments 
and crystals, respectively. These two- and three-soliton bound states are an example of 
the so-called soliton molecules. Recently, an optical equivalent of such objects has been 
realized experimentally in optical fibers~\cite{Stratmann:2005,Mitschke:2008} and 
a variety of theoretical proposals to create atomic soliton molecules have been 
presented~\cite{Kivshar:2003,Michinel:2011,Stoof:2011}. Soliton dimers share some properties 
with molecular dimers. However, as we discuss in detail below, intra-soliton interactions 
(of course absent in the case of individual polar molecules) are decisive for their stability 
and elementary excitations. Moreover, whereas molecular trimers may be found~(in absence 
of any additional lattice~\cite{Dalmonte:2011}) only for a rather narrow window of the dipole 
moment orientations~\cite{Wunsch:2011}, soliton trimers may exist for the orientations for 
which trimers of individual polar molecules are precluded.

The article is structured as follows. Sec.~\ref{sec:Model} introduces the general formalism. 
In Sec.~\ref{sec:Single} we compute the universal stability diagram for a single dipolar soliton 
in a quasi-1D trap and we show that such geometry supports two stable soliton regimes 
differing substantially in the character of the dipolar interactions. Section~\ref{sec:Dimer} is 
devoted to the study of properties of the soliton dimers. We discuss the inter-soliton binding 
potential and the nontrivial dependence of the dimer elementary excitations on the dipolar coupling. 
In Sec.~\ref{sec:Trimer} we analyze the trimer case, showing that soliton trimers may be found in 
a regime where molecular trimers would be unstable. We conclude in section~\ref{sec:Conclusions}.

\section{Model}
\label{sec:Model}

\begin{figure}[t]
\includegraphics[width=1.\columnwidth]{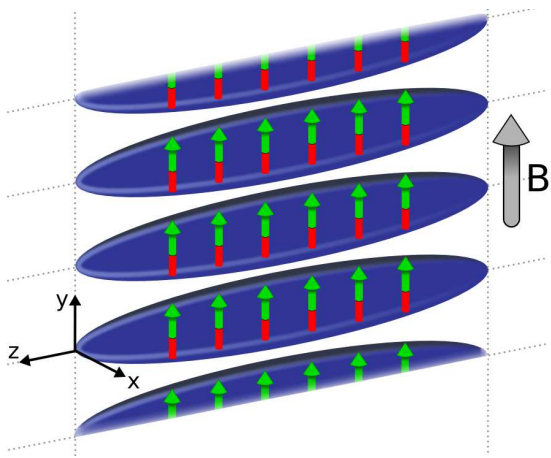}
\caption{(Color online) Scheme of a stack of quasi-1D 
tubes of dipolar Bose-Einstein condensates.}
\label{fig:1}
\end{figure}

In the following we consider a dipolar BEC loaded in a stack of $M$ parallel quasi-1D traps 
(tubes), formed by a 2D optical lattice with sites located at $y_j=j \Delta$ 
(Fig.~\ref{fig:1}).
The inter-tube potential barrier is considered sufficiently large to suppress any hopping. 
In each tube we assume a strong harmonic confinement of frequency $\omega_{\scriptscriptstyle \perp}$ 
in the $xy$ plane and no confinement along the $z$ direction. The atoms possess a magnetic 
dipole moment $\mu$ (the results are equally valid for electric dipoles, such as polar 
molecules) oriented along the $y$ axis, in the side-by-side configuration, by a sufficiently 
large external field. Introducing a wavefunction $\Psi_{j} \left( {\mbf{r}} \right)$ that 
describes an atomic cloud in a site $j$ holding $N$ atoms, the system of non-local coupled 
Gross-Pitaevskii equations (GPE) reads
\begin{multline}
\imath \hbar \partial_{t} \Psi_{j} \left( {\mbf{r}} \right) = \Biggl[
-\frac{\hbar^2}{2m}\nabla^{2} + U_{j}\left( {\mbf{r}} \right) 
  + g N \left| \Psi_{j} \left( {\mbf{r}} \right) \right|^{2} \\ 
  + \sum_{m=0}^{M-1} \int d \mbf{r'} V_{d} \left( \mbf{r} - \mbf{r'} \right) 
  \left| \Psi_{m} \left( {\mbf{r'}} \right) \right|^{2}
\Biggr] \Psi_{j} \left( {\mbf{r}}, t \right).
\label{eqn:coupledGP} 
\end{multline}
Here, $U_{j}\left( {\mbf{r}} \right) = \frac{1}{2}m \omega_{\scriptscriptstyle \perp}^{2} 
[ x^2 + \left( y- y_j \right)^2 ]$ and $V_{d}\left({\mbf{r}} - \mbf{r'}\right) = g_{d} N 
\left( 1 - 3 \cos^{2} \theta \right)/ \left| \mbf{r} - \mbf{r'}\right|^{3}$ is the dipole-dipole 
potential where $g_d=\mu_0 \mu^2/4 \pi$ with $\mu_{0}$ being the vacuum permeability and 
$\theta$ the angle between the vectors joining two interacting particles and the direction of 
the dipole moment. The short-range interactions are characterized by $g=4 \pi a_{sc} \hbar^2/m$ 
with $a_{sc}$ being the $s$-wave scattering length. In the following we consider attractive 
short-range interactions ($a_{sc}<0$).

\section{Dipolar soliton in a single quasi-1D trap}
\label{sec:Single}

\begin{figure}[b]
\includegraphics[width=1.\columnwidth]{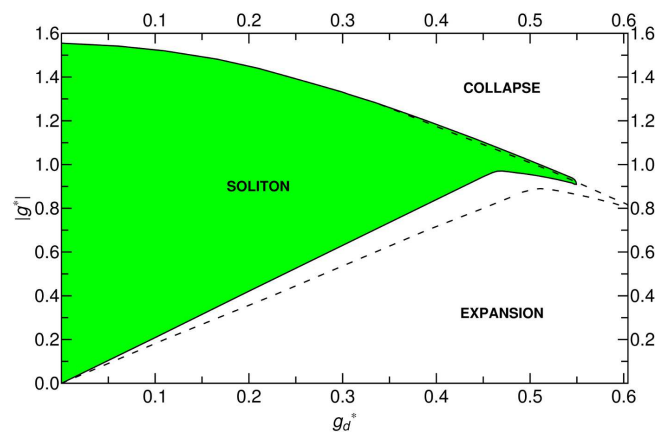}
\caption{(Color online) Universal stability diagram for a 
dipolar bright soliton in a single quasi-1D trap. Three 
regimes occur: stable soliton, instability against a 3D 
collapse and soliton expansion along the axis of the trap. 
The dashed line represents the stability boundary for a 
soliton dimer with $\Delta=6l_{\scriptscriptstyle \perp}$.}
\label{fig:2}
\end{figure}

We discuss first the conditions of existence of a stable bright soliton in a single 
quasi-1D trap~($M=1$). To this end we assume a 3D anisotropic Gaussian ansatz
\begin{equation}
 \Psi_{0} \left( {\mbf{r}} \right) = \frac{1}{\pi^{3/4} \left( l_x l_y l_z \right)^{1/2} } 
 \exp \left( -\frac{x^2}{2 l_{x}^{2}} -\frac{y^2}{2 l_{y}^{2}} -\frac{z^2}{2 l_{z}^{2}} \right),
\label{eqn:GaussianAnsatz}
\end{equation}
where $l_x$, $l_y$, and $l_z$ are the variational widths along $x$, $y$ and $z$ 
directions, respectively. Employing this ansatz into Eq.~\eqref{eqn:coupledGP} we obtain 
the energy of the system
\begin{multline}
 E\left( l_x,l_y,l_z \right) = \frac{\hbar^2}{4m} \sum_{i=x,y,z} \frac{1}{l_{i}^{2}} + 
 \frac{\omega_{\scriptscriptstyle \perp}^{2}}{4} \sum_{i=x,y}l_{i}^{2} \\ 
 + \frac{N}{4 \sqrt{2} \pi^{3/2} l_x l_y l_z} \left( g + \frac{2}{3} g_d \, 
   K \!\! \left( \frac{l_z}{l_x}, \frac{l_z}{l_y} \right) \right),
\label{eqn:EFunctional} 
\end{multline}
with the function
\begin{multline}
K \! \left(r_{x}, r_{y}\right) \\ \! = 
\int\limits_{0}^{2 \pi} \!\!d\varphi \!\! \int\limits_{0}^{1} \!\!du 
\frac{\left( 1- u^2 \right) \left[ 2 r_{y}^{2} - \left( r_{x}^{2} + 2 r_{y}^{2} \right) \cos^2 \varphi \right] - u^2}
{\left( 1- u^2 \right) \left[ r_{y}^{2} + \left( r_{x}^{2} - r_{y}^{2} \right) \cos^2 \varphi \right] + u^2}
\end{multline}
that in the cases of our interest may be evaluated analytically in terms of 
elliptic integrals~\cite{Gradshteyn:2000}. A stable soliton solution corresponds 
to a minimum in the energy functional $E\left( l_x,l_y,l_z \right)$ at finite non-zero 
values of the soliton widths. In Fig.~\ref{fig:2} we present the universal stability 
diagram as a function of the dimensionless parameters 
$g^{*} = gN/2 \pi \hbar \omega_{\scriptscriptstyle \perp} l_{\scriptscriptstyle \perp}^3$, 
and $g_{d}^{*} = g_{d} N / 2 \pi \hbar \omega_{\scriptscriptstyle \perp} l_{\scriptscriptstyle \perp}^3$.

\begin{figure}[t]
\includegraphics[width=1.\columnwidth]{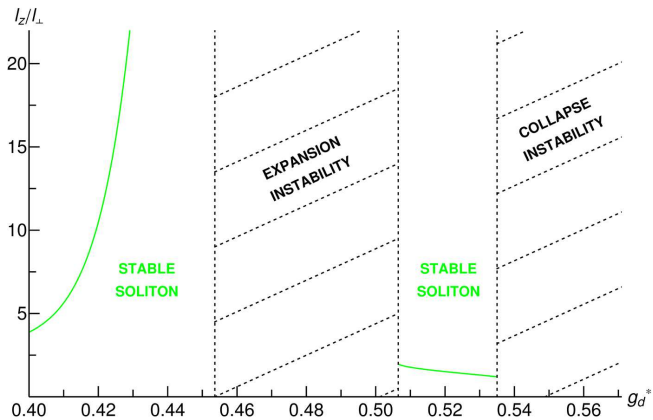}
\caption{(Color online) Reentrant character of the soliton 
stability in a single quasi-1D trap in the vicinity of 
$g^{*}_{c}$. Here, $|g^{*}|=0.95$.}
\label{fig:3}
\end{figure}

Interestingly, two different soliton regimes may be found, which differ 
remarkably in their properties and stability for growing $g_d>0$. For 
sufficiently small $|g^*|<|g_c^*|$, with $|g_c^*|\simeq 1$, a soliton 
may be considered as purely 1D, i.e. $l_x=l_y\simeq l_{\scriptscriptstyle \perp}
=\sqrt{\hbar/m\omega_{\scriptscriptstyle \perp}}$, whereas $l_z\gg l_{\scriptscriptstyle \perp}$. 
For such soliton, the DDI remains repulsive for any 
$g_d^*$. As a result, the soliton width $l_z$ increases monotonically for 
growing $g_d^*$, until diverging at a critical value at which the soliton 
delocalizes. The condition for soliton stability against the expansion may 
be then found analytically from Eq.~\eqref{eqn:EFunctional}, $|g^{*}|/g_d^{*}>2 \pi/3$ 
(straight solid line in Fig.~\ref{fig:1}). On the contrary, for $|g^*|>|g_c^*|$ 
the atomic cloud cannot be considered any more as 1D, since $l_z$ becomes 
comparable with the transversal widths. As a result, a stable soliton solution 
occurs that clearly displays a 3D character. In this regime, the DDI interaction 
changes its character from repulsive to attractive at a finite $g_d^*>0$ value, 
and hence for further growing $g_d^*$ the soliton width decreases until the 
soliton becomes unstable against 3D collapse. Furthermore, we note that in the vicinity 
of $|g_c^*|$, the stability diagram presents an interesting reentrant character 
as a function of $g_d$, first expanding, then re-binding and finally collapsing 
(Fig.~\ref{fig:3}). Interestingly, contrary to the soliton-expansion transition, 
at which the soliton width smoothly diverges, the re-binding transition is 
first-order-like, since the soliton abruptly re-binds at a finite width.

\section{Soliton dimers}
\label{sec:Dimer}

\begin{figure}[b]
\includegraphics[width=1.\columnwidth]{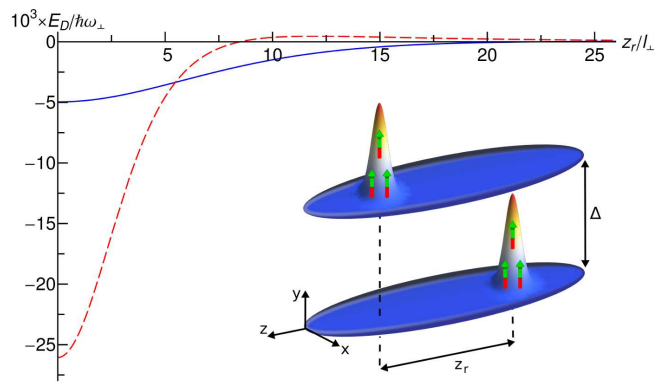}
\caption{(Color online) Inter-soliton binding potential 
for the case of the soliton dimer. The red dashed line represents the 
potential calculated within the point-like approximation $E_{D}^{0}$. 
The blue solid line shows the actual potential computed numerically 
with Eq.~\eqref{eqn:BindingEn}. Here, we consider the case of $^{52}$Cr 
condensate ($\mu \!=\! 6\, \mu_{B}$, with $\mu_{B}$ the Bohr magneton), 
$a_{sc}=-7.1\,a_{0}$ (with $a_0$ the Bohr radius), $N=100$, $\Delta = 
6 l_{\scriptscriptstyle \perp} = 512$ nm, and the lattice potential depth 
$s=13.3 E_{R}$ (recoil energy). These parameters refer to 
$\omega_{\scriptscriptstyle \perp} = 26.7$ kHz and $(g^{*},g_{d}^{*})=(-0.88,0.45)$. 
The inset depicts schematically the soliton dimer arrangement.}
\label{fig:4}
\end{figure}

We assume in the following that a soliton in each tube is in the 1D regime discussed 
in Sec.~\ref{sec:Single}~(this condition is self-consistently verified). At the end 
of this section we briefly comment on the case of solitons in the 3D regime. 
In the 1D regime, the wavefunctions factorize $\Psi_{j} \left( {\mbf{r}} \right) = 
\phi^{\scriptscriptstyle \perp}_{j} \left( x,y \right) \psi_{j} \left( z \right)$, 
with $\phi^{\scriptscriptstyle \perp}_{j}\left( x,y \right)$ the ground state wave 
function of the transverse harmonic oscillator in a site $j$. Employing the convolution 
theorem~\cite{Goral:2002} and integrating Eq. \eqref{eqn:coupledGP} with respect to 
$x$ and $y$, we arrive at the dimensionally reduced system of equations
\begin{multline}
 \imath \hbar \partial_{t} \psi_{j} \left( z \right) = \Biggl[ -\frac{\hbar^2}{2m} \partial_{z}^{2} 
 + \frac{g N}{2 \pi l_{\scriptscriptstyle \perp}^{2}} n_{j}(z) \\ 
 + \frac{g_d N}{3}\sum_{m=0}^{M-1}\! \int\! d k_z e^{\imath k_z z}  \hat{n}_{m} \left( k_z \right) 
 \!F_{m-j}\left(k_z \right) \! \Biggr] \psi_{j} \left( z \right),
\label{eqn:1DGPsystem}
\end{multline}
with $\hat{n}_{m} \left( k_z \right)$ the Fourier transform of the axial wave function 
density $n_{m}(z)=\left| \psi_{m} \left( z \right) \right|^{2}$ in a site $m$ and 
\begin{align}
F_{q} \left( k_z \right) = \! \int & \!\frac{d k_x d k_y}{\pi} \, 
\left( \frac{3 k_{y}^{2}}{ k_{x}^{2} + k_{y}^{2} + k_{z}^{2} } - 1 \right) \nonumber \\
& \times e^{-\frac{1}{2} \left( k_x^2 + k_y^2 \right) l_{\scriptscriptstyle \perp}^{2} - \imath k_y q \Delta }.
\end{align}
For stable individual solitons the inter-site DDI may result for $g_d>0$ 
in a binding of two solitons in different quasi-1D tubes into a soliton dimer 
(Fig.~\ref{fig:4}). This dimer resembles the case of the recently reported 
dimers of individual polar molecules. However, as discussed below, the interplay 
between intra-soliton interactions and inter-soliton interactions leads to a non-trivial 
effects in the physics of the soliton dimer, which do not occur in the case of molecular 
dimers due to the absence of on-site DDI.

Two solitons localized in neighboring quasi-1D tubes ($j=0,1$) 
and with a relative displacement $z_r$ along the axis direction 
$z$~(Fig.~\ref{fig:4}), experience an interaction potential 
\begin{align}
  \!\!\!E_{D}(z_r)\! =\! \frac{g_d N}{3} \!\! \int \!\! d z \, 
  n_{1} (z \!- \!z_r )\!\! \int \! dk_z e^{\imath k_z z}  \hat{n}_{0}\!\left( k_z \right) F_{1}(k_z).
\label{eqn:BindingEn}
\end{align}
We calculate $E_{D}(z_r)$ evolving Eq.~\eqref{eqn:1DGPsystem} in imaginary 
time to obtain the ground state of the dimer $\psi_j^{0}(z)$ and 
then shifting the solitons to the distance $z_{r}$. Due to the 
anisotropy of the DDI the inter-soliton potential is maximally 
attractive for $z_r=0$, becoming repulsive for large $z_r$ (Fig.~\ref{fig:4}). 
Naturally, the binding potential $E_{D}(z_r)$ calculated for actual 
soliton wave-packets is significantly weaker than that expected for 
point-like particles 
$E^{0}_{D}(z_r) = g_{d} N (z_{r}^2 - 2\Delta ^2)/(z_{r}^2+\Delta ^2)^{5/2}$. 
Nevertheless, we note that even for the case of the relatively small dipole 
moment of $^{52}$Cr, which we employed in our calculations for Fig.~\ref{fig:4}, 
the energy scale of the binding remains significant ($\sim 100$ Hz). 
The binding would be of course stronger for condensates of atoms with 
larger dipole moment, such as dysprosium~\cite{Lev:2011} and 
erbium~\cite{Ferlaino:2012}, or in the case of polar molecules 
\cite{Weidemuller:2008,Ni:2008,Ospelkaus:2010}. 

\begin{figure}[t]
\includegraphics[width=1.\columnwidth]{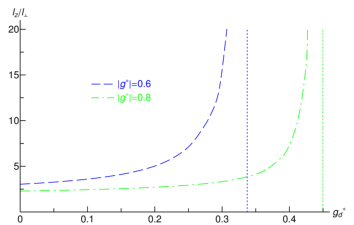}
\includegraphics[width=1.\columnwidth]{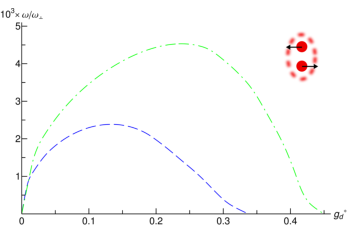}
\caption{(Color online) (top) Width of the soliton dimer as 
a function of $g_d^*$ for $|g^*|=0.6$ (blue dashed line) 
and $|g^*|=0.8$ (green dot-dashed line), for the same 
parameters as in Fig.~\ref{fig:4}. The vertical dashed lines 
indicate the dimer expansion threshold. (bottom) Frequency of 
elementary excitations of the soliton dimer for the same 
parameters. The inset shows a scheme of the dimer elementary 
excitation mode.}
\label{fig:5}
\end{figure}

We now focus on the essential properties of the soliton dimer. 
First, following the imaginary time evolution of Eq.~\eqref{eqn:1DGPsystem}, 
for a given $\Delta/l_{\scriptscriptstyle \perp}$, we compute the width $l_{z}$ of the solitons 
forming the dimer as a function of $g^{*}$ and $g_{d}^{*}$ 
(see Fig.~\ref{fig:5} (top)). Since we consider the 1D soliton regime, 
with an overall repulsive intra-soliton DDI, an increase of $g_{d}$ 
results in a broadening of the solitons, and eventually to the 
instability of the individual solitons against expansion. Note, however, 
that the attractive inter-soliton interactions, while providing the 
binding mechanism itself, induces a trapping of each soliton by its 
neighbor, which contributes to stabilization of each soliton against 
expansion. This increases the stability threshold found in Sec.~\ref{sec:Single} 
for an individual soliton, as shown by the straight dashed line 
(at $|g^{*}|/g_d^{*}=1.78$) in Fig.~\ref{fig:2}, obtained from a similar 
3D variational calculation as that of the previous section.

The properties of the soliton dimer must be compared with those 
of inter-site dimers formed by individual polar molecules. In the latter case, 
the localization of each molecular wave-packet is solely due to the 
attractive inter-site DDI, which induce a mutual trapping of both molecules. 
This means, in particular, that for $g_{d}^{*}=0$ each of the wave-packets 
delocalizes. Furthermore, owing to the absence of intra-wave-packet repulsive 
DDI, an increase of $g_{d}^{*}$ can only amplify localization and so the 
molecular dimer width decreases monotonically as a function of $g_{d}^{*}$, 
unlike the case of the soliton dimer. As a result, molecular dimers become 
stiffer (i.e. present growing excitation energies) for growing DDI.

On the contrary, the lowest-lying excitation of the soliton dimer 
presents a more involved behavior due to the interplay between 
intra- and inter-soliton DDI. We study the lowest-lying excitations 
by monitoring the real-time dynamics of the solitons following a small 
distortion of the ground state solution in the form 
$\psi_j(x,t=0)=\psi_j^{(0)}e^{-i (k_j x+\beta_j x^2)}$, corresponding to 
a perturbation of the soliton positions and their widths. Fig.~\ref{fig:5} 
(bottom) shows the result of the Fourier transform of the position $\aver{z(t)}$ 
of one of the two oscillating solitons and hence the frequency of the 
dimer lowest-lying excitation~(this is verified additionally by 
inspecting the Fourier transforms of soliton width and density oscillations). 
For sufficiently small DDI, and so for a small solitons widths, the 
lowest-lying excited mode of the dimer is associated exclusively to 
the motion of the center-of-mass of each soliton. In consequence, 
as $g_{d}^{*}$ grows, so does the energy of dimer excitations, 
resembling the case of molecular dimers. In contrast to the 
molecular dimers, however, after reaching a certain critical value 
of $g_{d}^{*}$ the soliton dimer becomes progressively softer (i.e. 
it exhibits decreasing excitation energies). This phenomenon arises because 
the soliton widths increase due to the repulsive intra-soliton DDI, and 
as a result the lowest-lying excitation becomes eventually an admixture 
of both position and width distortions. As discussed before, for a sufficiently 
large $g_d^*$  the dimer becomes eventually unstable against expansion. 

Finally, we stress that soliton dimers may exist as well in 
the 3D regime defined in Sec.~\ref{sec:Single}, i.e. for $|g^*|>|g_c^*|$. 
As depicted in Fig.~\ref{fig:2}, the stability threshold against the 
soliton dimer collapse is basically the same as that for an individual 
soliton. Contrary to the 1D case, in the 3D regime the width of a soliton 
is relatively small $l_z \simeq l_{\scriptscriptstyle \perp}$ and so the binding 
potential between the two solitons, such as the one depicted in Fig.~\ref{fig:4}, 
becomes comparably deeper, approaching the point-like approximation $E_{D}^{0}$. 
Moreover, for $|g^*|>|g_c^*|$ the soliton width never becomes large enough to 
cause the mixing of position and width excitations. As a result, in the 3D regime, 
for growing $g_d$ values the soliton dimer becomes only stiffer, up to the 
collapse threshold, similary to the case of molecular dimers.

\section{Soliton trimers}
\label{sec:Trimer}

\begin{figure}[b]
\includegraphics[width=1.\columnwidth]{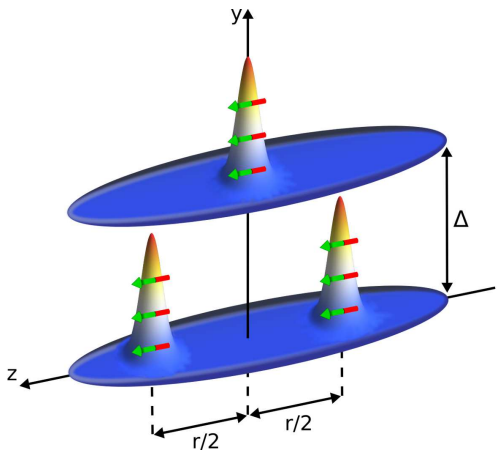}
\caption{Scheme of the soliton trimer. The dipole moments 
are aligned in the head-to-tail configuration 
providing attractive intra-site and repulsive 
inter-site dipolar interactions. In our work we mimick 
this scheme with qualitatively equivalent arrangement 
of dipoles aligned along the $y$ axis (side-by-side 
configuration) but with $g_{d}^{*}<0$ (see text).}
\label{fig:6}
\end{figure}
\begin{figure}[t]
\includegraphics[width=1.\columnwidth]{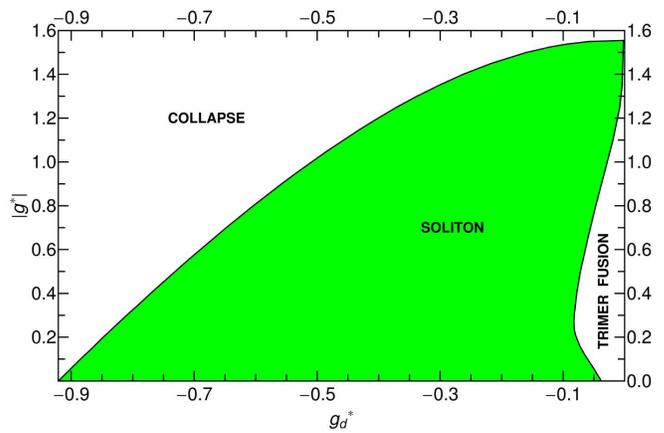}
\caption{(Color online) Universal stability diagram of 
a single dipolar bright soliton in a quasi-1D 
trap for $g_{d}^{*}<0$. Two regimes occur: stable 
soliton and 3D instability against collapse. 
We indicate additionally the regime of the 
trimer instability against fusion (see discussion 
in text).}
\label{fig:7}
\end{figure}

Interestingly, the DDI may lead to the formation of soliton molecules 
comprising of more than two solitons, in particular soliton trimers 
(Fig.~\ref{fig:6}). We note that trimers (and even more involved complexes) 
have been predicted as well for individual polar molecules~\cite{Wunsch:2011,
Dalmonte:2011}. However, molecular trimers have been found 
to exist only in a rather narrow window of dipole moment orientations with 
respect to the trap axis, in the very vicinity of the magic angle, 
such that intra-site repulsion is minimized and inter-site attraction 
is maximized. In particular, molecular trimers are precluded if 
the dipole orientation is aligned along the trap axis. Furthermore, 
as noted in Sec.~\ref{sec:Dimer}, the formation of molecular bound 
states is handicapped by the fact that the inter-site interactions 
do not only provide a binding between the molecules but are also 
indispensable for the localization of the individual molecular wave-packets 
themselves. This contrasts with the soliton case, where the existence 
of localized wavepackets is supported by intra-soliton interactions. 
As a result, as we discuss in this section, the interplay between 
inter- and intra-soliton interactions allows for stable soliton trimers 
for dipole moment orientations in which molecular trimers are absent. 

In the following we consider for theoretical simplicity the case 
of dipoles oriented along the $y$ axis (in the side-by-side configuration 
as that of the soliton dimer) but with $g_d<0$. This may be achieved by means 
of a rotating magnetic field~\cite{Giovanazzi:2002}, or microwave 
dressing for polar molecules~\cite{Micheli:2007}. The results would 
be however qualitatively very similar to the case of dipoles oriented along the 
tubes, since both cases are characterized by repulsive inter-site 
DDI and attractive intra-site DDI. Although the attractive inter-site 
interactions seem naively to involve soliton fusion in the bottom tube, 
and hence to preclude the existence of the soliton trimer, such trimer 
results actually from a nontrivial interplay between inter-tube repulsion 
and intra-tube attraction. Namely, the single soliton in the upper tube 
provides a repulsive potential barrier that prevents the fusion of the two 
mutually-attracting solitons in the bottom tube, hence keeping the soliton 
trimer stable. 

A major difference with respect to soliton dimers lies in the fact that 
now $g_d^*<0$, and hence the intra-soliton interaction is attractive. As a 
result for growing $|g_d^*|$ the individual solitons shrink, i.e. the trimer 
is not unstable against the expansion of the individual solitons but rather 
against their collapse, since the solitons become eventually 3D for a sufficiently 
large $|g_d^*|$. As shown in the dimer case, the threshold for the collapse instability 
is basically given by the intra-soliton physics. We have hence analyzed the 
stability of a soliton in a single quasi-1D trap for $g_d<0$~(see Fig.~\ref{fig:7}), 
using the same 3D variational Gaussian ansatz discussed in Sec.~\ref{sec:Single}. Naturally, 
soliton trimers may exist only within the stability region of individual solitons. 

\begin{figure}[b]
\includegraphics[width=1.\columnwidth]{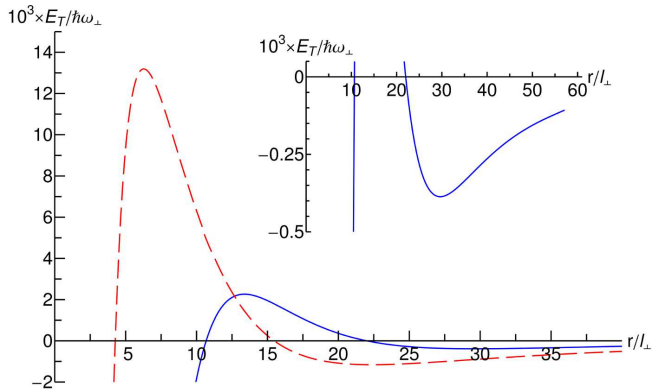}
\caption{(Color online) Soliton trimer potential energy. The red dashed 
line represents the potential $E_{T}^{0}$ obtained from the point-like 
approximation, whereas the blue solid line depicts the actual binding potential 
$E_{T}$ calculated numerically integrating Eqs.~\eqref{eqn:1DGPsystem}. 
We consider $^{52}$Cr BEC (with $g_d<0$) with $a_{sc}=-4.0\,a_{0}$ and $N=100$ 
atoms in every soliton, i.e. $(g^{*},g_{d}^{*})=(-0.50,-0.45)$. Here, 
$\Delta =512$ nm, $s=13.3E_{R}$ and $\omega_{\scriptscriptstyle \perp} = 26.7$ kHz. 
The inset shows the potential energy minimum which sustains the trimer 
bound state.}
\label{fig:8}
\end{figure}

\begin{figure}[t]
\includegraphics[width=1.\columnwidth]{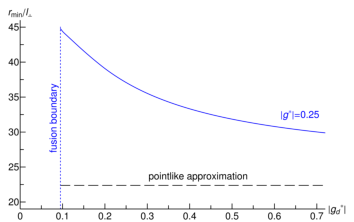}
\includegraphics[width=1.\columnwidth]{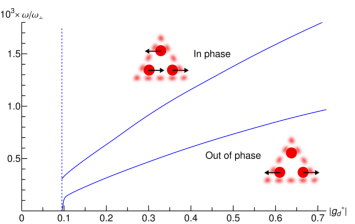}
\caption{(Color online) (top) Size of the soliton 
trimer as a function of $g_d^*$ for $|g^*|=0.25$ and 
the remaining lattice parameters as that in the Fig.~\ref{fig:8}. 
The dashed line indicates the result from the point-like approximation 
$E_{T}^{0}$. (bottom) Trimer elementary excitations for the same parameters. 
The insets show schemes of the trimer elementary excitation modes.}
\label{fig:9}
\end{figure}

In the following we analyze the properties of trimers well within 
the 1D regime, i.e. far from the 3D collapse threshold, for which we 
can safely employ the 1D GPEs (Eq.~\eqref{eqn:1DGPsystem}). In particular, 
after obtaining the ground state of the trimer configuration by means of 
the imaginary time evolution of these equations, we have computed 
the binding potential of the trimer $E_{T}(r)$~(Fig.~\ref{fig:8}) 
as a function of the distance $r$ between the solitons in the bottom 
tube~(Fig.~\ref{fig:6}). Crucially, at an intermediate distance $r_{min}$ 
$E_{T}(r)$ shows a local minimum that offers the possibility of a soliton trimer. 
A point-like approximation of the solitons would induce a binding  
\begin{align}
  E^{0}_{T}(r) =g_d N \left [ \frac{1}{r^3}+\frac{16(r^2-8\Delta^2)}
    {(r^2+4\Delta^2)^{5/2}} \right ],
\label{eqn:SimplifiedTrimerPotential}
\end{align}
resulting in an equilibrium position $r^{0}_{min}/\Delta \simeq 3.73$, 
independently of $g_d$. This approximation, however, departs significantly from 
the actual binding potential $E_{T}(r)$, proving again the relevance of the 
spatial extension of solitons. Specifically, as shown in Fig.~\ref{fig:9} (top), 
the trimer size, understood as the actual equilibrium distance $r_{min}$, 
decreases with growing $|g_d^*|$~(whereas the binding energy increases). We also 
note that, as it may be expected, the soliton trimer is more loosely bound than 
the soliton dimer. For typical parameters of $^{52}$Cr condensate, which we 
employed for the Fig.~\ref{fig:8}, the binding energy is of the order of $10$Hz. 
We stress, however, that the binding will be certainly stronger in the case of 
more magnetic atoms (Dy, Er) or polar molecules.

Note that contrary to the dimer case, the soliton trimer is related with a local 
minimum of the energy functional. In particular, the global energy minimum results 
from the fusion of the solitons in the bottom tube into a single soliton, which forms 
a tilted dimer with the top soliton. The trimer configuration of Fig.~\ref{fig:6} 
is hence strictly speaking a metastable solution, which is separated from the fused 
solution by a potential barrier (Fig.~\ref{fig:8}). Macroscopic quantum
tunneling through this barrier is negligible, and hence the metastable solution may 
be considered for all practical purposes as stable (as we have checked in real-time 
evolution). The potential barrier dissappears at a sufficiently small $|g_d^*|$, 
at which the soliton trimer becomes abruptly unstable against soliton fusion (see 
Fig.~\ref{fig:7} and Fig.~\ref{fig:9}).

Finally, as for the case of the soliton dimers, we have analyzed the lowest-lying 
excitations of the soliton trimer. Since now $g_d<0$, the solitons are always well 
localized. Hence, contrary to the dimer case, the lowest lying excitations are related 
solely to the solitons center of mass motion (without an excitation of the width of 
the solitons). We may hence define two different types of elementary excitations,  
characterized by an in-phase and an out-of-phase motion of the soliton pair in 
the bottom tube, respectively (Fig.~\ref{fig:9}~(bottom)). As in Sec.~\ref{sec:Dimer}, 
we have probed these modes perturbing the soliton widths and positions of the trimer ground 
state, and monitoring the subsequent real-time dynamics given by Eq.~\eqref{eqn:1DGPsystem}. 
After Fourier transforming the soliton positions, we obtain the lowest-lying excitations as 
a function of $|g_{d}^{*}|$. The results of the two excitation frequencies are depicted in 
Fig.~\ref{fig:9}~(bottom), which shows that for all $|g_d^*|$ the out-of-phase mode is always 
less energetic than the in-phase mode. 

\section{Conclusions}
\label{sec:Conclusions}

In summary, inter-site dipolar interactions support the formation 
of soliton molecules in a stack of quasi-1D tubes. The stability properties of 
quasi-1D solitons and inter-site soliton molecules depend crucially on the 
interplay between dipolar and contact interactions, and the competition between 
intra-site and inter-site effects. In particular, two different quasi-1D soliton regimes 
are possible: 1D solitons for which the intra-soliton DDI is always repulsive and 
that become eventually unstable against soliton delocalization, and 3D solitons, for which 
the DDI changes its character from repulsive to attractive for growing DDI, and that become 
eventually unstable against soliton collapse. We have shown that, contrary to the case of 
dimers of individual polar molecules, the inter-play between intra-soliton interactions 
and inter-soliton DDI leads to a non-trivial behavior of the lowest-lying excitations of 
soliton dimers. In the purely-1D regime a growing DDI render the dimer stiffer up to a 
maximum beyond which an increasing DDI softens the dimer due to the admixture 
between position and width excitations. Finally, we have shown that soliton trimers may be 
constructed for attractive intra-site and repulsive inter-site DDI due to a subtle 
interplay between intra-tube attraction and inter-tube repulsion. Interestingly, these 
trimers occur in a regime in which trimers of individual polar molecules are not possible. 
The reported soliton molecules can be observed under realistic conditions within current 
experimental feasibilities. Moreover, we emphasize that the soliton binding mechanism described 
in this work can be straightforwardly generalized to engineer even more intricate soliton 
complexes comprising a larger number of solitons in more sites of an optical lattice.

\section{Acknowledgement}
\label{sec:acknow}

We thank M. Jona-Lasinio for fruitful discussions. We acknowledge funding by the 
German-Israeli Foundation, the Cluster of Excellence QUEST and the DFG~(SA1031/6). 

\bibliographystyle{apsrev4-1}
\bibliography{SolitonMolecules}{}

\end{document}